\documentclass[epj]{webofc}
\usepackage[varg]{txfonts}   
\usepackage{graphicx,color} 
\usepackage{bm}       
\usepackage{amsmath}
\usepackage{amssymb}
\newcommand{\ket}[1]{ | #1 \rangle}

\woctitle{21st International Conference on Few-Body Problems in Physics}
\begin{document}
\title{The Pion-Cloud Contribution to the\\ Electromagnetic Nucleon Form Factors}
\author{Daniel Kupelwieser\inst{1}\fnsep\thanks{\email{con-fuser43@web.de}} \and
        Wolfgang Schweiger\inst{1}\fnsep\thanks{\email{wolfgang.schweiger@uni-graz.at}}
}

\institute{Institute of Physics, University of Graz, A-8010 Graz, Austria
          }

\abstract{
We study the electromagnetic structure of the nucleon within a hybrid constituent-quark model that comprises, in addition to the $3q$ valence component, also a $3q$+$\pi$ non-valence component. To this aim we employ a Poincar\'e-invariant multichannel formulation based on the point-form of relativistic quantum mechanics. With a simple 3-quark wave function for the bare nucleon, i.e. the $3q$-component, we obtain reasonable results for the nucleon form factors and predict the meson-cloud contribution to be significant only below $Q^2\lesssim 0.5$~GeV$^2$ amounting to about 10\% for $Q^2\rightarrow 0$, in accordance with the findings of other authors.}
\maketitle
\section{Introduction}
Pions, undoubtedly, play an important role in explaining the nucleon structure~\cite{TW2001}. In the present contribution we try to substantiate the physical picture of a nucleon as consisting of a quark core that is surrounded by a pion cloud by means of a hybrid constituent-quark model in which a physical nucleon is a superposition of a $3q$ and a $3q$+$\pi$ component. Thereby both, the constituent quarks and and the pion, are treated as dynamical degrees of freedom. Within our model pions couple directly to the constituent quarks which are, in addition, subject to an instantaneous confining force. Such kind of models have also been considered by other authors~\cite{Miller:2002ig,Pasquini:2007iz}. These authors use the light-front formalism to account for relativity in a proper way, we rather adopt the point-form approach which has already turned out to be rather convenient for form-factor calculations of mesonic systems~\cite{Biernat:2009my,Biernat:2010tp,GomezRocha:2012zd,Biernat:2014dea}.

\section{Formalism}\label{sec:formalism}
Following the same strategy as in Refs.~\cite{Biernat:2009my,Biernat:2010tp,GomezRocha:2012zd,Biernat:2014dea} we calculate the invariant $1$$\gamma$-exchange electron-nucleon scattering amplitude, extract the electromagnetic nucleon current, analyze the covariant structure of the current and identify the electromagnetic nucleon form factors. In order to derive the $1$$\gamma$-exchange amplitude we employ a Poincar\'e invariant coupled-channel formulation which relies on the point form of relativistic quantum mechanics~\cite{Biernat:2010tp}. Typical for the point-form, all four components of the momentum operator are interaction dependent, whereas the generators of Lorentz transformations stay free of interactions.  As a welcome consequence one has simple rotation and boost properties and angular-momentum addition works like in non-relativistic quantum mechanics. Poincar\'e invariance is achieved by employing the Bakamjian-Thomas construction which allows to separate the overall motion of the system from the internal motion in a neat way:
\begin{equation}\label{eq:massop}
\hat{P}^{\mu}=\hat{\mathcal M}\,  \hat V^{\mu}_{\mathrm{free}}=
\left(\hat{\mathcal M}_{\mathrm{free}}+ \hat{\mathcal
M}_{\mathrm{int}} \right) \hat V^{\mu}_{\mathrm{free}}\, ,
\end{equation}
i.e. the 4-momentum operator factorizes into an interaction-dependent mass operator $\hat{\mathcal{M}}$ and a free 4-velocity operator $ \hat V^{\mu}_{\mathrm{free}}$. Bakamjian-Thomas-type mass operators are most conveniently represented in a velocity-state basis.
Velocity states $\vert V;  {\bf k}_1, \mu_1; {\bf k}_2, \mu_2; \dots ; {\bf k}_n, \mu_n\rangle$ are characterized by the overall velocity $V$ ($V_\mu V^\mu=1$) of the system, the CM momenta ${\bf k}_i$ of the individual particles and their (canonical) spin projections $\mu_i$~\cite{Biernat:2010tp}.

We now want to calculate the $1$$\gamma$-exchange amplitude for elastic electron scattering off a nucleon that consists of a $3q$ and a $3q$+$\pi$ component. Thereby not only the dynamics of electron and quarks, but also the dynamics of the photon and the pion should be fully taken into account. This is accomplished by means of a multichannel formulation that comprises all states which can occur during the scattering process (i.e. $|3q, e \rangle$, $|3q, \pi, e \rangle$, $|3q, e, \gamma \rangle$, $|3q, \pi, e, \gamma \rangle$). What one then needs, in principle, are scattering solutions of
\begin{equation}\label{EVequation}
\left(\begin{array}{cccc}
\hat{M}_{3qe}^{\mathrm{conf}} & \hat{K}_\pi & \hat{K}_\gamma & \hat{K}_{\pi\gamma}
\\
\hat{K}_\pi^\dagger & \hat{M}_{3q \pi e} ^{\mathrm{conf}}& \hat{\tilde K}_{\pi\gamma} &
\hat{K}_\gamma \\
\hat{K}_\gamma^\dagger & \hat{\tilde K}_{\pi\gamma}^\dag & \hat{M}_{3q e \gamma} ^{\mathrm{conf}}&
\hat{K}_\pi \\
\hat{K}_{\pi\gamma}^\dag & \hat{K}_\gamma^\dagger & \hat{K}_\pi^\dagger &
\hat{M}_{3q \pi e \gamma}^{\mathrm{conf}}
\end{array}\right)
\left(\begin{array}{l}
\ket{\psi_{3q e}} \\ \ket{\psi_{3q \pi e}} \\
\ket{\psi_{3q e \gamma}} \\ \ket{\psi_{3q \pi e \gamma}}
\end{array}\right)
=
\sqrt{s} \left(\begin{array}{l}
\ket{\psi_{3q e}} \\ \ket{\psi_{3q \pi e}} \\
\ket{\psi_{3q e \gamma}} \\ \ket{\psi_{3q \pi e \gamma}}
\end{array}\right)
\end{equation}
which evolve from an asymptotic electron-nucleon in-state $\ket{e N}$ with invariant mass $\sqrt{s}$. The diagonal entries of this matrix mass operator contain, in addition to the relativistic kinetic energies of the particles in the particular channel, an instantaneous confinement potential between the quarks. The off-diagonal entries are vertex operators which describe the transition between the channels. In the velocity-state representation these vertex operators are directly related to  usual quantum-field theoretical interaction-Lagrangean densities~\cite{Biernat:2010tp}. The 4-vertices $\hat{K}_{\pi\gamma}^{(\dag)}$ and $\hat{\tilde{K}}_{\pi\gamma}^{(\dag)}$ show up only for pseudovector pion-quark coupling and vanish for pseudoscalar pion-quark coupling which we will concentrate on in the following.

\begin{figure*}[!t]
\centering
\begin{minipage}{0.25\textwidth}
\vspace{-1.5em}
\includegraphics[width=0.9\textwidth]{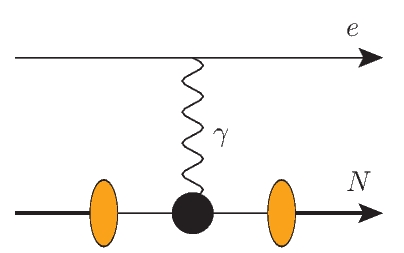}
\end{minipage}
\begin{minipage}{1ex}
\vspace{-1em}
+
\end{minipage}
\begin{minipage}{0.30\textwidth}
\includegraphics[width=0.9\textwidth]{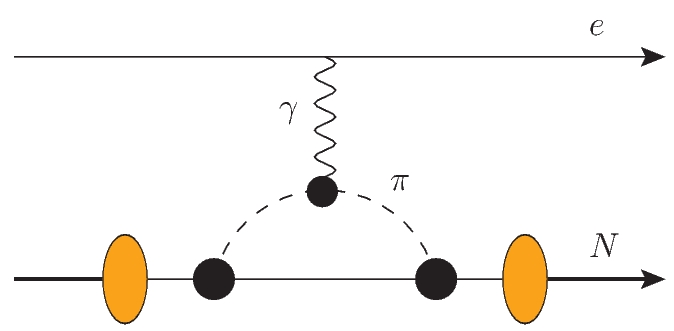}
\end{minipage}
\begin{minipage}{1ex}
\vspace{-1em}
+
\end{minipage}
\begin{minipage}{0.30\textwidth}
\vspace{-1.0em}
\includegraphics[width=0.9\textwidth]{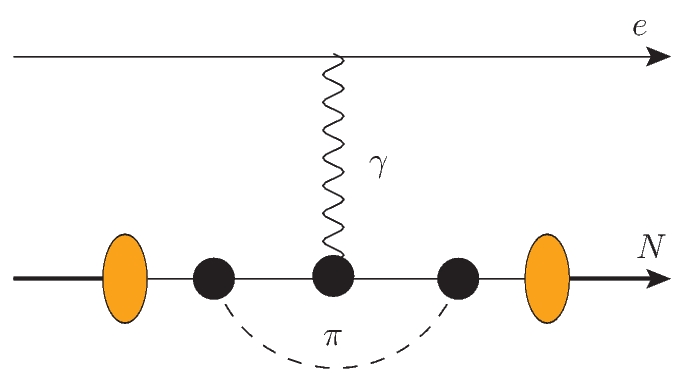}
\end{minipage}
\caption{Diagrams representing the $1$$\gamma$-exchange amplitude for electron scattering off a \lq\lq physical\rq\rq\ nucleon $N$, i.e. a bare nucleon dressed by a pion cloud. The time orderings of the $\gamma$-exchange are subsumed under a covariant photon propagator. Black blobs represent vertex form factors for the coupling of a photon or pion to the bare nucleon $\tilde N$. A vertex form factor is also assumed at the photon-pion vertex. The ovals represent the wave function (i.e. essentially the probability $P_{\tilde N/N}$) for finding the bare nucleon in the physical nucleon. \label{overallff}}
\end{figure*}

At this point it is convenient to reduce Eq.~(\ref{EVequation}) to an eigenvalue problem for $\ket{\psi_{3q e}} $ by means of a Feshbach reduction:
\begin{equation}\label{eq:Mphys}
\left[\hat{M}_{3qe} +\hat{K}_\pi(\sqrt{s}-\hat{M}_{3q\pi e} )^{-1} \hat{K}_\pi^\dag + \hat{V}_{1\gamma}^{\mathrm{opt}}(\sqrt{s})\right] \ket{\psi_{3q e}} = \sqrt{s} \, \ket{\psi_{3q \pi e}} \, ,
\end{equation}
where $\hat{V}_{1\gamma}^{\mathrm{opt}}(\sqrt{s})$ is the 1$\gamma$-exchange optical potential. The invariant 1$\gamma$-exchange electron-nucleon scattering amplitude is now obtained by sandwiching $\hat{V}_{1\gamma}^{\mathrm{opt}}(\sqrt{s})$ between (the valence component of) physical electron-nucleon states  $\ket{eN}$, i.e. eigenstates of $[  \hat{M}_{3qe} +\hat{K}(\sqrt{s}-\hat{M}_{3q\pi e} )^{-1} \hat{K}^\dag ]$. The crucial point is now to observe that, due to instantaneous confinement, propagating intermediate states do not contain free quarks, they rather contain either physical nucleons  $N$ or bare baryons $\tilde{B}$, the latter being eigenstates of the pure confinement problem. This allows us to rewrite the scattering amplitude in terms of pure hadronic degrees of freedom with the quark substructure being hidden in vertex form factors. This is graphically represented in Fig.~\ref{overallff}.

In a first step only nucleons are taken into account in the pion loop. To calculate the diagrams shown in Fig.~\ref{overallff} we thus need the electromagnetic and strong form factors of the bare nucleon and the the electromagnetic form factor of the pion. The microscopic expression (in terms of quark-degrees of freedom) for the first diagram in Fig.~\ref{overallff} gives the electromagnetic form factors of the bare nucleon.\footnote{For details of their extraction and a discussion of the problems connected with wrong cluster properties associated with the Bakamjian-Thomas construction we refer to Refs.~\cite{Biernat:2009my,Biernat:2010tp,Biernat:2014dea}.}  What enters into the analytical expressions for these form factors is the 3-$q$ bound-state wave function of the bare nucleon. Instead of solving the bound-state problem for a particular confinement potential we rather make an ansatz for this wave function. For comparison with corresponding front-form calculations~\cite{Miller:2002ig,Pasquini:2007iz} the 3-quark (momentum space) bound-state wave function of the bare nucleon, which we also need to calculate the strong $\pi\tilde{N}\tilde{N}$ form factor, is taken as
\begin{equation}\label{eq:nucleonwf}
\Phi_{\tilde N}\left({\tilde {\bf k}}_i\right)=\frac{\mathcal{N}}{\left((\sum \tilde{\omega}_i)^2 + \beta^2\right)^\gamma}\, ,
\end{equation}
with ${\tilde {\bf k}}_i$ and $\tilde{\omega}_i$ denoting the quark momenta and energies in the rest frame of the nucleon and the parameters $\beta$, $\gamma$ taken from Ref.~\cite{Pasquini:2007iz}. The normalization $\mathcal{N}$ has to be fixed such that the whole nucleon wave function, including the $3q$+$\pi$ component, is normalized to one. The electromagnetic pion form factor is taken from Ref.~\cite{Biernat:2009my}, where it has been calculated within the same approach as here using a harmonic-oscillator model for the $u\bar d$ bound-state wave function of the $\pi^+$. An analogous calculation gives also the strong coupling and the strong form factor for the $\pi\tilde{N}\tilde{N}$ vertex~\cite{Kupelwieser:2013nqa}.
Here we want to emphasize that, unlike the authors of Refs.~\cite{Miller:2002ig,Pasquini:2007iz}, who took a phenomenological $\pi\tilde{N}\tilde{N}$ form factor, we have calculated both, the electromagnetic form factors of the bare nucleon as well as the strong $\pi\tilde{N}\tilde{N}$ vertex form factor with the same microscopic input, namely the 3-quark bound-state wave function $\Phi_{\tilde{N}}$ given in Eq.~(\ref{eq:nucleonwf}).

\section{Results and Outlook}\label{sec:results}

\begin{figure}
\centering
\includegraphics[width=0.48\textwidth]{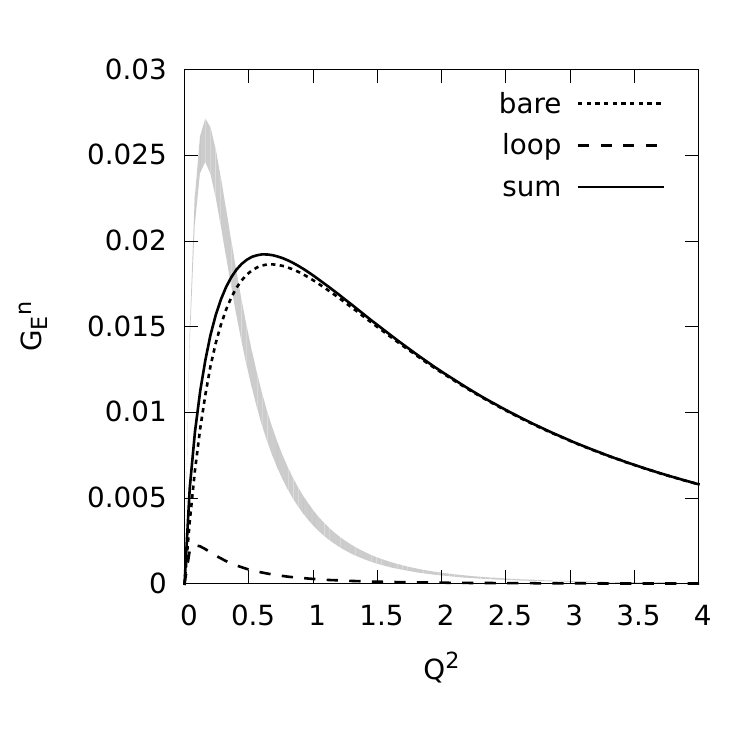}
\includegraphics[width=0.48\textwidth]{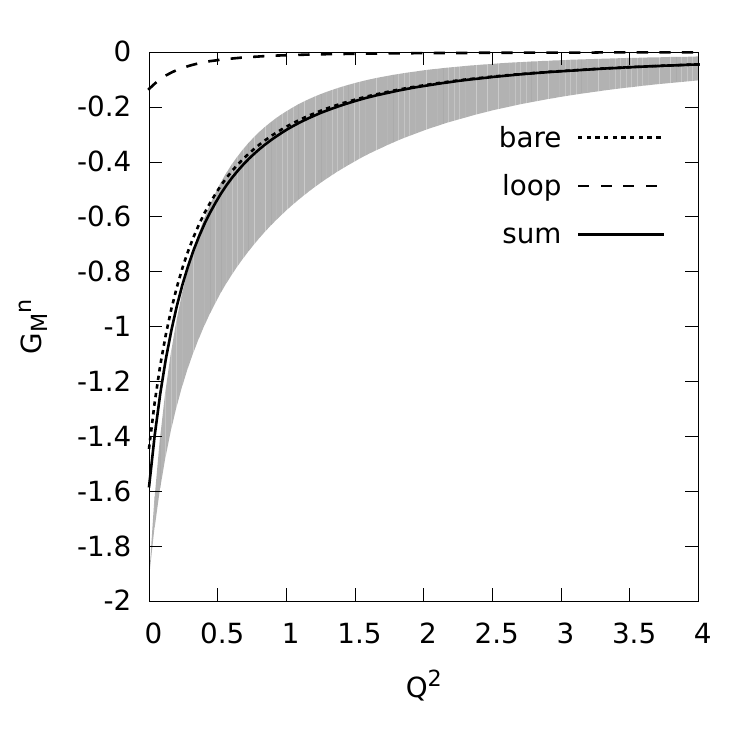}\vspace{-0.7cm}
\caption{The neutron electric (left) and magnetic (right) form factors as predicted by our model (solid line). The $3q$ valence contribution is indicated by the dotted line, the contribution due to the $3q+\pi$ non-valence component by the dashed line. The shaded area is a parameterization of the experimental data (including uncertainties)~\cite{Kelly:2004hm}.} \label{fig:neutronffs}
\end{figure}

With the model sketched above we achieve good agreement with the experimental data for proton electric and magnetic form factors. The corresponding results for the neutron electric and magnetic form factors are shown in Fig.~\ref{fig:neutronffs}.  Whereas our neutron magnetic form factor is also in reasonable agreement with a parameterization of the corresponding experimental data, the reproduction of the neutron electric form factor seems to be less satisfactory. But here one has to notice that it is a rather small quantity and the error bars on the experimental data points are, in general, much larger than the indicated uncertainty of the parameterization. The  size of the pionic contribution to all the nucleon form factors is comparable with the one found in Refs.~\cite{Pasquini:2007iz,Miller:2002ig}. A significant effect of the $3q+\pi$ component on the form factors is only observed for  momentum transfers $Q^2\lesssim 0.5$~GeV$^2$, where it leads to a welcome modification of the $Q^2$-dependence.
%

To improve the model one may first think of a more sophisticated 3-$q$ wave function of the (bare) nucleon, containing, e.g., a mixed $S\!U(6)$ spin-flavor-symmetry component like in Ref.~\cite{Pasquini:2007iz}. Further improvements concern the replacement of the pseudoscalar by the pseudovector $\pi \tilde{N}\tilde{N}$ coupling, which guarantees correct properties in the chiral limit. This would entail additional pion-loop graphs with a $\gamma\pi \tilde{N}\tilde{N}$ vertex. A complete treatment of the pion-loop contribution will also require to take into account other baryons, different from the nucleon, within the loop, the lightest and most important of them being the $\Delta$.  We do not expect that this kind of refinements, which are the subject of ongoing work, will drastically change the size of the pion-loop contribution, but they hopefully will help to achieve an even better reproduction of the electromagnetic nucleon form factors.

\begin{acknowledgement}
D. Kupelwieser acknowledges the support of the \lq\lq Fonds zur
F\"orderung der wissenschaftlichen Forschung in \"Osterreich\rq\rq\ (FWF
DK W1203-N16).
\end{acknowledgement}

%

\begin{thebibliography}{99}

\bibitem{TW2001}
A.~W.~Thomas and W.~Weise, {\em The Structure of the Nucleon} (Wiley-VCH, Berlin, 2001).

\bibitem{Miller:2002ig}
  G.~A.~Miller,
  Phys.\ Rev.\ C {\bf 66}, 032201 (2002).

\bibitem{Pasquini:2007iz}
  B.~Pasquini and S.~Boffi,
  Phys.\ Rev.\ D {\bf 76},  074011 (2007).

\bibitem{Biernat:2009my}
  E.~P.~Biernat, W.~Schweiger, K.~Fuchsberger and W.~H.~Klink,
  Phys.\ Rev.\ C {\bf 79}, 055203 (2009).

\bibitem{Biernat:2010tp}
  E.~P.~Biernat, W.~H.~Klink and W.~Schweiger,
  Few Body Syst.\  {\bf 49}, 149 (2011).

\bibitem{GomezRocha:2012zd}
  M.~Gomez-Rocha and W.~Schweiger,
  Phys.\ Rev.\ D {\bf 86}, 053010 (2012).

\bibitem{Biernat:2014dea}
  E.~P.~Biernat and W.~Schweiger,
  Phys.\ Rev.\ C {\bf 89}, 055205 (2014).

\bibitem{Kupelwieser:2013nqa}
D.~Kupelwieser and W.~Schweiger,
  Few Body Syst.\  {\bf 55}, 881 (2014).



\bibitem{Kelly:2004hm}
  J.~J.~Kelly,
  Phys.\ Rev.\ C {\bf 70}, 068202 (2004).

\end{thebibliography}
%
%

\end{document}